# Gain-assisted superluminal microwave pulse propagation via four-wave mixing in superconducting phase quantum circuits

Z. Amini Sabegh, A. Vafafard, M. A. Maleki, and M. Mahmoudi*

Department of Physics, University of Zanjan, University Blvd, 45371-38791, Zanjan, Iran

Tel. : +98-24-23052521

Fax: +980241-23052264

*E-mail: mahmoudi@znu.ac.ir

**Abstract**

We study the propagation and amplification of a microwave field in a four-level cascade quantum system which is realized in a superconducting phase quantum circuit. It is shown that by increasing the microwave pump tones feeding the system, the normal dispersion switches to the anomalous and the gain-assisted superluminal microwave propagation is obtained in this system. Moreover, it is demonstrated that the stimulated microwave field is generated via four-wave mixing without any inversion population in the energy levels of the system (amplification without inversion) and the group velocity of the generated pulse can be controlled by the external oscillating magnetic fluxes. We also show that in some special set of parameters, the absorption-free superluminal generated microwave propagation is obtained in superconducting phase quantum circuit system.

PACS numbers: 42.50.Gy, 03.67.Lx, 74.50.+r, 85.25.Cp

## Introduction

It is well known that a superconducting quantum circuit (SQC), containing the Josephson junctions, has a set of quantized energy levels and behaves as an artificial atom. It has been used to establish a number of the atomic physics and the quantum optics phenomena [1]. A Josephson junction is a weak link between two superconducting bulks which is separated by an insulating layer and acts as a device with nonlinear inductance and no energy dissipation [2]. The main differences between the SQC and a natural atom lie on the energy difference scale of their energy levels and the system coupling with its environment. The energy scale in the SQC corresponds to the gigahertz and the coupling in SQC is also stronger than the natural atoms. Moreover, there are not usually well-defined selection rules for interaction of superconducting circuits with the microwave fields [3].

The superconducting charge, flux and phase quantum circuits based on the Josephson junctions are macroscopic in size but they show quantum mechanical behaviors such as having discrete energy levels, superposition of states and entanglement. Then, it have been extensively used as a powerful tool in quantum information processing [4- 5]. The generation, conversion, amplification and propagation of the microwave signals are important because of their potential applications in measuring and controlling the qubits in solid state quantum information processing.

The atomic system has been employed to establish the several interesting nonlinear coherent optical phenomena, i.e. electromagnetically induced transparency (EIT) [6], Autler-Townes splitting [7], coherent population trapping [8], amplification without inversion [9], wave mixing [10], slow light [11] and fast light [12]. Such behaviors are not restricted to the atomic systems and the solid state systems can be also used to establish the quantum phenomena. The EIT-like phenomenon has been introduced in superconducting systems in rotating wave approximation [13] and beyond it [14]. The phase-sensitive properties of the closed-loop systems, established in the SQC, have been studied in $\Delta$-type configuration and it was shown that EIT window can be controlled by relative phase of driving fields [15]. The Autler-Townes doublet has been experimentally measured in a superconducting charge [16] and phase qubits [17]. The EIT, decoherence, Autler-Townes and dark states have been theoretically investigated in two-tone driving of a three-level superconducting phase [18] and flux quantum circuit [19]. Direct observation of coherent population trapping has been reported in the $\Lambda$-type configuration of superconducting phase qubits [20]. Coherent microwave pulse control of quantum memory via slow light in SQC was studied and realization of the coherent storage and on-demand pulse retrieval are reported in this system [21]. The difference- and sum-frequency generation via three-wave mixing was investigated in the microwave regime using a single three-level superconducting flux quantum circuit [4]. The steady state and dynamical behavior of a $V$-type artificial atom, established in a superconducting quantum interference device, were studied and it was shown that the ratio of the Josephson coupling energy to the capacitive coupling strength has a significant impact on determining of the dynamical behavior and creating the optical bistability [22-23]. Controlling of the group velocity in a SQC was investigated and superluminal pulse propagation was reported in the presence of the absorption peak in the absorption spectrum [24] but, because of the attenuation, it is difficult to propagate the pulse in the presence of such absorption peak.

In this paper, we investigate the propagation and amplification without inversion of microwave pulse in a superconducting phase quantum circuit (SPQC). We show that by changing the applied time-dependent magnetic flux amplitudes, the slope of dispersion can be switched from positive to negative which corresponds to the superluminal pulse propagation. Moreover, such propagation happens in a transparent window and it is accompanied by the gain doublet without appearance of any inversion population in energy levels. In the next step, we investigate the microwave generated pulse in the SPQC via four-wave mixing. The four-wave mixing has been investigated in atomic systems [25, 26]. We find that the stimulated microwave field is generated without any inversion of population and propagates in superluminal region. Then stimulated generation of superluminal microwave pulse propagation is obtained in the SPQC via four-wave mixing process.

## Model and equations

We introduce a SPQC consisting of a Josephson junction in a superconducting circuit with a gate capacitance $C$, inductance $L$ and Josephson junction inductance $L_J$ which is placed in an external magnetic flux as a driving element. The schematic of the system is shown in Fig. 1(a). The external magnetic flux consists of a constant term $\Phi_{dc}$ and an oscillating term $\Phi_{rf}(t)$, so that $\Phi_{ext} = \Phi_{dc} + \Phi_{rf}(t)$. The Josephson junction energy is denoted by $E_J = \Phi_0^2 / (4\pi^2 L_J)$, where $\Phi_0 = 2.07\times10^{-15} Wb$ is quantum of the flux. Using the critical Josephson current, $I_c = (2\pi/\Phi_0)E_J$, the Josephson junction potential energy is given by $E_J(1-\cos(2\pi\Phi/\Phi_0))$ where $\Phi$ stands for the flux variable. Then the classical Hamiltonian of the SPQC can be written as

$$H = \frac{Q^2}{2C} + \frac{(\Phi-\Phi_{ext})^2}{2L} - E_J \cos(2\pi\frac{\Phi}{\Phi_0}). \tag{1}$$

By assuming $|\Phi_{rf}(t)| << |\Phi_{dc}|$ the Hamiltonian switches to

$$H = \frac{Q^2}{2C} + \frac{(\Phi-\Phi_{dc})^2}{2L} - E_J \cos(2\pi\frac{\Phi}{\Phi_0}) - \frac{\Phi\Phi_{rf}(t)}{L}, \tag{2}$$

where $2\pi\Phi/\Phi_0$ stands for superconducting phase difference across the junction. The first term, in the right hand side of Eq. (2) is the circuit kinetic energy and two next terms are the circuit potential energy. The last term is time dependent part of the classical Hamiltonian. Then the potential energy of the circuit can be written as

$$U(\varphi) = U_0\left[(\varphi-\varphi_{dc})^2 - \beta\cos(2\pi\varphi)\right], \tag{3}$$

where $U_0 = \Phi_0^2/(2L)$, $\beta = L/(2\pi^2 L_J)$, $\varphi = \Phi/\Phi_0$ and $\varphi_{dc} = \Phi_{dc}/\Phi_0$. Note that for $\varphi_{dc} = 0.5$ the potential energy mentioned in Eq. (3), is symmetric around $\varphi = 0.5$.

In Fig. 1(b), we plot the potential energy for $L = 2L_J = 700\,pH$ and $\Phi_{dc} = 0.54\Phi_0$ versus $\varphi$ between zero and unity. Noting that two minima of the potential energy in this interval are close to $\varphi = 0.22$ and $\varphi = 0.82$, their locations are approximately given by

$$\varphi_{m1} = 0.22 + \frac{\varphi_{dc} - \pi\beta\sin(0.44\pi) - 0.22}{1 + 2\pi^2\beta\cos(0.44\pi)},$$

$$\varphi_{m2} = 0.75 - \frac{1}{4\pi^3\beta} + \sqrt{\frac{1}{2\pi^2} + \frac{1}{2\pi^3\beta}\left(\varphi_{dc} - \frac{3}{4}\right) + \frac{1}{16\pi^6\beta^2}}. \tag{4}$$

The condition for existence of the second minimum is

$$\varphi_{dc} > \frac{3}{4} - \pi\beta - \frac{1}{8\pi^3\beta}. \tag{5}$$

Now, we choose the first minimum, $\varphi_{m1}$, and expand the potential energy as

$$\begin{aligned} \frac{U(\Delta\varphi)}{U_0} &= (\varphi_{m1} - \varphi_{dc})^2 - \beta\cos(0.44\pi) + 2\pi\beta\sin(0.44\pi)(\varphi_{m1} - 0.22) \\ &+ \left[1 + 2\pi^2\beta\cos(0.44\pi) - 4\pi^3\beta\sin(0.44\pi)(\varphi_{m1} - 0.22)\right](\Delta\varphi)^2 \\ &- \frac{4\pi^3\beta}{3}\left[\sin(0.44\pi) + 2\pi\cos(0.44\pi)(\varphi_{m1} - 0.22)\right](\Delta\varphi)^3, \end{aligned} \tag{6}$$

where $\Delta\varphi = \varphi - \varphi_{m1}$. Dropping the first constant term, shifting the reference point to the first minimum $\varphi_{m1}$ and neglecting the third term in the vicinity of $\varphi = \varphi_{m1}$, we obtain

$$U(\Phi) = \frac{\Phi^2}{2L_J^*}, \tag{7}$$

where $L_J^* = L_J[(2\pi^2\beta) + \cos(0.44\pi) - 2\pi\sin(0.44\pi)(\varphi_{m1} - 0.22)]^{-1}$.Then the time independent part of the classical Hamiltonian becomes as

$$H_0 = \frac{Q^2}{2C} + \frac{\Phi^2}{2L_J^*}. \tag{8}$$

To quantize the Hamiltonian of Eq. (8), we consider $\Phi$ and $Q$ as the canonical conjugate variables which satisfy the commutation relation $[\Phi, Q] = i\hbar$. The creation ($a^\dagger$) and annihilation ($a$) operators for the harmonic oscillator can be written as

$$a = \sqrt{\frac{C\omega_0}{2\hbar}}\left(\Phi + \frac{iQ}{C\omega_0}\right), \qquad a^\dagger = \sqrt{\frac{C\omega_0}{2\hbar}}\left(\Phi - \frac{iQ}{C\omega_0}\right), \tag{9}$$

where $\omega_0 = 1/\sqrt{L_J^* C}$ is the Josephson plasma frequency. The operators $a$ and $a^\dagger$ obey the commutation relation $[a, a^\dagger] = 1$ and $H_0$ reads to

$$H_0 = \hbar\omega_0\left(a^\dagger a + 1/2\right), \tag{10}$$

with the eigenvalue equation $H_0|n\rangle_0 = \hbar\omega_0(n+1/2)|n\rangle_0$. We can shift all the energy levels in order to get the zero ground state energy.

The energy levels of this Hamiltonian are the same as the case of the harmonic oscillator. In order to construct an artificial atomic system interacting with microwave fields, we add the time dependent part, $(-\Phi/L)\Phi_{rf}(t)$ as an interaction term to the Hamiltonian. Then the Hamiltonian matrix of the four-level artificial atom interacting with microwave fields is given by

$$H = H_0 - \frac{\Phi\Phi_{rf}(t)}{L} = \hbar\begin{bmatrix} 0 & \frac{1}{\sqrt{2}}g(t) & 0 & 0 \\ \frac{1}{\sqrt{2}}g(t) & \bar{\omega}_{10} & g(t) & 0 \\ 0 & g(t) & \bar{\omega}_{10}+\bar{\omega}_{21} & \sqrt{\frac{3}{2}}g(t) \\ 0 & 0 & \sqrt{\frac{3}{2}}g(t) & \bar{\omega}_{10}+\bar{\omega}_{21}+\bar{\omega}_{32} \end{bmatrix}, \tag{11}$$

where $\bar{\omega}_{10}$, $\bar{\omega}_{21}$ and $\bar{\omega}_{32}$ are the central frequencies of corresponding transitions. Moreover the coupling parameter is defined by

$$g(t) = -\frac{\Phi_{rf}(t)}{L\sqrt{\hbar C\omega_0}}. \tag{12}$$

Equation (11) introduces the Hamiltonian of a ladder type artificial atomic system interacting with microwave fields. We consider the time-dependent part of the external flux as

$$\Phi_{rf}(t) = \Phi_{10}\cos(\omega_{10}t) + \Phi_{21}\cos(\omega_{21}t) + \Phi_{32}\cos(\omega_{32}t), \tag{13}$$

which can excite the three transitions $|0\rangle \to |1\rangle$, $|1\rangle \to |2\rangle$ and $|2\rangle \to |3\rangle$ as shown in Fig. 1(b). The parameters $\omega_{10}$, $\omega_{21}$ and $\omega_{32}$ are the frequencies of the applied external microwave fields. Thus, the coupling parameter is given by

$$g(t) = g_{10}\cos(\omega_{10}t) + g_{21}\cos(\omega_{21}t) + g_{32}\cos(\omega_{32}t), \tag{14}$$

where the applied time-dependent flux amplitudes are

$$g_{10} = -\frac{\Phi_{10}}{L\sqrt{\hbar C\omega_0}}, \quad g_{21} = -\frac{\Phi_{21}}{L\sqrt{\hbar C\omega_0}}, \quad g_{32} = -\frac{\Phi_{32}}{L\sqrt{\hbar C\omega_0}}. \tag{15}$$

Note that in the presence of the third term of Eq. (6), i.e. for strong amplitudes of the external time-dependent terms, our SPQC model will not be equivalent as a four-level cascade quantum system.

We are going to use the density matrix formalism to investigate the dynamical properties of the SPQC system. Then we should construct the density matrix in the interaction picture as $\rho = U^{\dagger}\rho_S U$ where $\rho_S$ is the density matrix in the Schrödinger's picture. The following unitary matrix

$$U = \begin{bmatrix} 1 & 0 & 0 & 0 \\ 0 & e^{-i\omega_{10}t} & 0 & 0 \\ 0 & 0 & e^{-i(\omega_{10}+\omega_{21})t} & 0 \\ 0 & 0 & 0 & e^{-i(\omega_{10}+\omega_{21}+\omega_{32})t} \end{bmatrix}, \tag{16}$$

transforms the operators from Schrödinger's to the interaction picture. Then we get

$$\rho = \begin{bmatrix} \tilde{\rho}_{00} & \tilde{\rho}_{01} & \tilde{\rho}_{02} & \tilde{\rho}_{03} \\ \tilde{\rho}_{10} & \tilde{\rho}_{11} & \tilde{\rho}_{12} & \tilde{\rho}_{13} \\ \tilde{\rho}_{20} & \tilde{\rho}_{21} & \tilde{\rho}_{22} & \tilde{\rho}_{23} \\ \tilde{\rho}_{30} & \tilde{\rho}_{31} & \tilde{\rho}_{32} & \tilde{\rho}_{33} \end{bmatrix}, \tag{17}$$

where the matrix elements are given by

$$\begin{aligned} &\tilde{\rho}_{00} = \rho_{00},\ \tilde{\rho}_{01} = e^{-i\omega_{10}t}\rho_{01},\ \tilde{\rho}_{02} = e^{-i(\omega_{10}+\omega_{21})t}\rho_{02},\ \tilde{\rho}_{03} = e^{-i(\omega_{10}+\omega_{21}+\omega_{32})t}\rho_{03}, \tilde{\rho}_{11} = \rho_{11}, \\ &\tilde{\rho}_{12} = e^{-i\omega_{21}t}\rho_{12},\ \tilde{\rho}_{13} = e^{-i(\omega_{21}+\omega_{32})t}\rho_{13},\ \tilde{\rho}_{22} = \rho_{22},\ \tilde{\rho}_{33} = \rho_{33}. \end{aligned} \tag{18}$$

The Hamiltonian in the interaction picture can be also followed by

$$\tilde{H} = U^{\dagger}HU + i\hbar\left(\frac{\partial U^{\dagger}}{\partial t}\right)U = \hbar\begin{bmatrix} 0 & \frac{\Omega_{10}}{2} & 0 & 0 \\ \frac{\Omega_{10}}{2} & -\Delta_{10} & \frac{\Omega_{21}}{2} & 0 \\ 0 & \frac{\Omega_{21}}{2} & -(\Delta_{10}+\Delta_{21}) & \frac{\Omega_{32}}{2} \\ 0 & 0 & \frac{\Omega_{32}}{2} & -(\Delta_{10}+\Delta_{21}+\Delta_{32}) \end{bmatrix}, \tag{19}$$

where $\Omega_{10} = \frac{1}{\sqrt{2}}g_{10}$, $\Omega_{21} = g_{21}$ and $\Omega_{32} = \sqrt{\frac{3}{2}}g_{32}$ are the Rabi frequencies of applied fields. The detuning of applied fields with corresponding transitions are denoted by
$\Delta_{10} = \omega_{10} - \overline{\omega}_{10}$, $\Delta_{21} = \omega_{21} - \overline{\omega}_{21}$, $\Delta_{32} = \omega_{32} - \overline{\omega}_{32}$.

The master equation for density matrix operator in the interaction picture is

$$\frac{\partial\rho}{\partial t} = -\frac{i}{\hbar}\left[\tilde{H},\rho\right] + \tilde{L}[\rho], \tag{20}$$

where the operator $\tilde{L}[\rho]$, indicating the decaying between the quantum states, can be written as [27, 28]

$$\tilde{L}[\rho]=\sum_{i\in\{1,2,3\}}\frac{\Gamma_{i,i-1}}{2}\left(2\sigma_{i-1,i}\rho\sigma_{i,i-1}-\sigma_{ii}\rho-\rho\sigma_{ii}\right)-\sum_{i,j\in\{0,1,2,3\};i\neq j}\frac{\gamma_{ij}}{2}\sigma_{ii}\rho\sigma_{jj}\,. \qquad (21)$$

Here $\gamma_{ij}=\gamma_{ji}$ is the pure inter-level dephasing rate and $\sigma_{ij}=|i\rangle\langle j|$. The parameter $\Gamma_{ij}$ is the inter-level relaxation rate of the transition $|i\rangle\to|j\rangle$. Then the equations of motion for the density matrix elements are given by

$$\dot{\tilde{\rho}}_{00}=-i\frac{\Omega_{10}}{2}(\tilde{\rho}_{10}-\tilde{\rho}_{01})+\Gamma_{10}\tilde{\rho}_{11},$$

$$\dot{\tilde{\rho}}_{01}=-i\left[\frac{\Omega_{10}}{2}(\tilde{\rho}_{11}-\tilde{\rho}_{00})+\Delta_{10}\tilde{\rho}_{01}-\frac{\Omega_{21}}{2}\tilde{\rho}_{02}\right]-\frac{1}{2}(\Gamma_{10}+\gamma_{10})\tilde{\rho}_{01},$$

$$\dot{\tilde{\rho}}_{02}=-i\left[\frac{\Omega_{10}}{2}\tilde{\rho}_{12}+(\Delta_{10}+\Delta_{21})\tilde{\rho}_{02}-\frac{\Omega_{21}}{2}\tilde{\rho}_{01}-\frac{\Omega_{32}}{2}\tilde{\rho}_{03}\right]-\frac{1}{2}(\Gamma_{21}+\gamma_{20})\tilde{\rho}_{02},$$

$$\dot{\tilde{\rho}}_{03}=-i\left[\frac{\Omega_{10}}{2}\tilde{\rho}_{13}+(\Delta_{10}+\Delta_{21}+\Delta_{32})\tilde{\rho}_{03}-\frac{\Omega_{32}}{2}\tilde{\rho}_{02}\right]-\frac{1}{2}(\Gamma_{32}+\gamma_{30})\tilde{\rho}_{03},$$

$$\dot{\tilde{\rho}}_{11}=-i\left[\frac{\Omega_{10}}{2}(\tilde{\rho}_{01}-\tilde{\rho}_{10})+\frac{\Omega_{21}}{2}(\tilde{\rho}_{21}-\tilde{\rho}_{12})\right]-\Gamma_{10}\tilde{\rho}_{11}+\Gamma_{21}\tilde{\rho}_{22},$$

$$\dot{\tilde{\rho}}_{12}=-i\left[\frac{\Omega_{21}}{2}(\tilde{\rho}_{22}-\tilde{\rho}_{11})+\Delta_{21}\tilde{\rho}_{12}+\frac{\Omega_{10}}{2}\tilde{\rho}_{02}-\frac{\Omega_{32}}{2}\tilde{\rho}_{13}\right]-\frac{1}{2}(\Gamma_{10}+\Gamma_{21}+\gamma_{21})\tilde{\rho}_{12},$$

$$\dot{\tilde{\rho}}_{13}=-i\left[\frac{\Omega_{10}}{2}\tilde{\rho}_{03}+(\Delta_{21}+\Delta_{32})\tilde{\rho}_{13}+\frac{\Omega_{21}}{2}\tilde{\rho}_{23}-\frac{\Omega_{32}}{2}\tilde{\rho}_{12}\right]-\frac{1}{2}(\Gamma_{10}+\Gamma_{32}+\gamma_{31})\tilde{\rho}_{13},$$

$$\dot{\tilde{\rho}}_{22}=-i\left[\frac{\Omega_{32}}{2}(\tilde{\rho}_{32}-\tilde{\rho}_{23})+\frac{\Omega_{21}}{2}(\tilde{\rho}_{12}-\tilde{\rho}_{21})\right]+\Gamma_{32}\tilde{\rho}_{33}-\Gamma_{21}\tilde{\rho}_{22},$$

$$\dot{\tilde{\rho}}_{23}=-i\left[\frac{\Omega_{32}}{2}(\tilde{\rho}_{33}-\tilde{\rho}_{22})+\Delta_{32}\tilde{\rho}_{23}+\frac{\Omega_{21}}{2}\tilde{\rho}_{13}\right]-\frac{1}{2}(\Gamma_{21}+\Gamma_{32}+\gamma_{32})\tilde{\rho}_{23},$$

$$\tilde{\rho}_{00}+\tilde{\rho}_{11}+\tilde{\rho}_{22}+\tilde{\rho}_{33}=1\,. \qquad (22)$$

The SPQC absorption and amplification properties as well as the group velocity of microwave field can be obtained from the linear susceptibility, $\chi=F\,|\,d_{10}\,|^2\,/(V\varepsilon_0\hbar\Omega_{10})\rho_{10}$, which is the response of the system to the applied fields. Here $F$, $d_{10}$, $\varepsilon_0$ and $V$ stand for the optical confinement factor, the dipole moment vector, the vacuum permittivity and the volume of the single SPQC, respectively. Real and imaginary parts of the susceptibility are considered as the dispersion and absorption responses of the system. In our notation, the positive value of the imaginary part corresponds to the absorption, while the negative value shows the gain.

In the special conditions $\Delta_{10} << \Gamma$, $\gamma_{ij} = 0$, $\Gamma_{10} = \Gamma_{32} = \Gamma$, $\Gamma_{21} = 2\Gamma$ and for the weak probe field, $\Omega_{10} << \Gamma$, the following analytical expression is obtained for the probe transition coherence

$$\rho_{10} = \frac{2\Omega_{10}[4\Gamma^4 - \Gamma^2(\Omega_{21}^2 - 4\Omega_{32}^2) + \Omega_{32}^2(\Omega_{21}^2 + \Omega_{32}^2)]}{\Gamma^2(2\Gamma^2 + \Omega_{21}^2 + \Omega_{32}^2)^2}\Delta_{10} - \frac{i(2\Gamma^2 + \Omega_{32}^2)\Omega_{10}}{\Gamma^2(2\Gamma^2 + \Omega_{21}^2 + \Omega_{32}^2)^2} . \quad (23)$$

According to Eq. (23) the slope of dispersion can be switch from positive to negative for small probe detuning. The imaginary part is negative for all value of the Rabi frequencies which is corresponding to a gain in SPQC. Such behavior is similar to the four-level cascade type atomic system.

In a dispersive medium the different frequency components of a pulse will experience different refractive index and then each frequency component in the pulse travels at different velocity. The group velocity of a pulse, velocity of the pulse peak, in a dispersive medium is determined by the slope of dispersion. We introduce the group index $n_g = c / v_g$ where the group velocity $v_g$ at the frequency $\omega$ is given by

$$v_g = \frac{c}{1 + \frac{1}{2}\chi'(\omega) + \frac{\omega}{2}\frac{\partial\chi'(\omega)}{\partial\omega}} = \frac{c}{n_g} . \quad (24)$$

According to Eq. (24), the group velocity of the pulse in a dispersive medium can exceed the velocity of light in vacuum ($c$), leading to the superluminal pulse propagation. It is worth nothing that the group velocity can be different from the information velocity and then such propagation does not violate the special relativity principle of Einstein. In our notation the negative slope of dispersion corresponds to the anomalous dispersion, while the positive slope shows the normal dispersion.

**Results and discussions**

Now, we are interested in summarize the numerical results of Eqs. (22). We consider a weak probe field which is applied to the transition $|0\rangle \to |1\rangle$ and investigate the absorption and propagation of the probe field. We scale all frequency parameters by $\Gamma_{10}$. Figure 2 shows the real (a) and imaginary (b) parts of the susceptibility versus probe detuning for different values of the Rabi frequency of applied microwave fields. Used parameters are $\Gamma_{10} = \Gamma = 14\pi\, MHz$, $\Gamma_{21} = 1.57\Gamma$, $\Gamma_{32} = \Gamma$, $\gamma_{10} = \Gamma$, $\gamma_{20} = 2.29\Gamma$, $\gamma_{21} = 2.57\Gamma$, $\gamma_{30} = 2.43\Gamma$, $\gamma_{31} = 1.71\Gamma$, $\gamma_{32} = \Gamma$, [18] and the fields parameters are $\Delta_{32} = 0$, $\Delta_{21} = 0$, $\Omega_{10} = 0.01\Gamma$, $\Omega_{32} = \Gamma$, $\Omega_{21} = \Gamma$ (solid), $5\Gamma$ (dashed), $8\Gamma$ (dash-dotted). An investigation on Fig. 2 shows that for $\Omega_{21} = \Gamma$ the slope of dispersion around zero probe detuning is positive and it accompanies by a gain dip. But by increasing the Rabi frequency of the applied microwave fields the slope of dispersion switches to the negative corresponding to the superluminal pulse propagation. It is worth to note that such propagation is accompanying by a gain doublet with negligible value around zero probe detuning. Then the gain-assisted superluminal pulse propagation is obtained in the SPQC system. The numerical results are in good agreement with Eq. (23). The corresponding group index behaviors versus probe detuning are shown in Fig. 3. Used parameters are same as in Fig. 2. It is clear that for small values of the Rabi frequencies the group index around $\Delta_{10} = 0,$ is greater than unity which means the group velocity is less than $c$. But by increasing the Rabi frequency the group index becomes less than unity or even negative leading to the group velocity greater than $c$ or negative group velocity, respectively.

The amplification without inversion is another scenario which we are following in SPQC system. Our analytical result in Eq. (23) shows that the probe field will amplify for all values of the Rabi frequency of applied fields. We are going to investigate the inversion in the population of energy levels. The population difference of probe transition versus Rabi frequencies is plotted in Fig. 4. Used parameters are same as in Fig. 2. It is found that the population inversion does not happen and amplification without inversion is established in this system.

In the next step, we are interested in investigating the four-wave mixing in SPQC system. We apply three microwave fields to the corresponding transitions as shown in Fig. 1(b). By annihilation of three photons, a fourth microwave photon with frequency $\omega_{FWM} = \omega_{10} + \omega_{21} + \omega_{32}$ is generated via four-wave mixing. We focus on the third-order susceptibility medium response in the frequency $\omega_{FWM}$ which is corresponding to $\rho_{30}$, the coherence term of transition $|0\rangle \leftrightarrow |3\rangle$. In Fig. 5, we plot the real (a) and imaginary (b) parts of $\rho_{30}$ versus the $\Delta = \Delta_{10} + \Delta_{21} + \Delta_{32}$, for $\Delta_{21} = \Delta_{32} = 0,$ $\Omega_{10} = \Omega_{21} = \Omega_{32} = 2.1\Gamma$ (solid), $5\Gamma$ (dashed). Other parameters are same as in Fig. 2. We find that the system shows the gain at the frequency $\omega_{FWM}$ via four-wave mixing process. Moreover the slope of dispersion for stimulated generated pulse becomes negative via the changing microwave pump tones feeding the system, so that for small values of the Rabi frequencies the slope of dispersion around zero detuning is negative. An investigation on the absorption spectrum shows that the generated superluminal pulse at the central frequency $\omega_{FWM}$ is accompanied by a gain doublet. Then the gain-assisted superluminal pulse propagation is obtained for stimulated generated microwave field via four-wave mixing process in the SPQC system.

Note that the structure and dips position of the doublet gain can understand via the four dressed states of the SPQC system which can be written as

$$\lambda_1 = -\lambda_2 = \frac{\sqrt{2}}{4}\left(\sqrt{\Omega_{10}{}^2 + \Omega_{21}{}^2 + \Omega_{32}{}^2 + \sqrt{\Omega_{10}{}^4 + 2\Omega_{10}{}^2(\Omega_{21}{}^2 - \Omega_{32}{}^2) + (\Omega_{21}{}^2 + \Omega_{32}{}^2)^2}}\right),$$

$$\lambda_3 = -\lambda_4 = \frac{\sqrt{2}}{4}\left(\sqrt{\Omega_{10}{}^2 + \Omega_{21}{}^2 + \Omega_{32}{}^2 - \sqrt{\Omega_{10}{}^4 + 2\Omega_{10}{}^2(\Omega_{21}{}^2 - \Omega_{32}{}^2) + (\Omega_{21}{}^2 + \Omega_{32}{}^2)^2}}\right). \quad (25)$$

Each of two dips in the gain doublet are approximately located at $\Delta = \lambda_1$ and $\Delta = \lambda_2$, however they are slightly shifted by including the pure inter-level dephasing rates.

Finally we are looking for the amplification without inversion of the generated four-wave mixing pulse. In Fig. 6 we plot the population difference states $|0\rangle$ and $|3\rangle$ versus the Rabi frequencies for zero detuning of the applied fields. Used parameters are same as in Fig. 2. It is clear that the inversion of population does not appear in the transition $|0\rangle \leftrightarrow |3\rangle$ and stimulated generation of superluminal pulse propagation happens without inversion of population in SPQC system.

The population difference of transition $|0\rangle \leftrightarrow |3\rangle$ for the parameters $\gamma_{ij} = 0$, $\Gamma_{10} = \Gamma_{32} = \Gamma$, $\Gamma_{21} = 2\Gamma$, $\Omega_{10} = \Omega_{21} = \Omega_{32} = \Omega$, $\Delta_{10} = \Delta_{21} = \Delta_{32} = 0$, is given by

$$\rho_{33} - \rho_{00} = -\frac{72\Gamma^8 + 264\Gamma^6\Omega^2 + 268\Gamma^4\Omega^4 + 65\Gamma^2\Omega^6}{72\Gamma^8 + 336\Gamma^6\Omega^2 + 442\Gamma^4\Omega^4 + 258\Gamma^2\Omega^6 + 40\Omega^8}. \quad (26)$$

This expression reveals that for the conditions mentioned above, the population inversion cannot establish in transition $|0\rangle \leftrightarrow |3\rangle$.

## Conclusion

We have investigated the generation and propagation of the microwave pulse in a four-level SPQC system. We found that by increasing the microwave pump tones feeding the system, the normal dispersion switches to the anomalous accompanied by the gain doublet. It was shown that the stimulated generated pulse via four-wave mixing propagates in superluminal condition and can be controlled by the external oscillating magnetic fluxes. Moreover, we found that the generation of superluminal microwave pulse appears without any inversion of the population of the transition states. The establishing of the absorption-free superluminal microwave pulse propagation and amplification of microwave field without inversion as well as the stimulated superluminal pulse propagation via four-wave mixing can be used to control of the SQCs and to quantum information transfer in these systems.

## Figures Caption

**Figure 1.** (a) Schematic of SPQC consisting of a Josephson junction in a superconducting circuit with a gate capacitance $C$, inductance $L$ and Josephson junction inductance $L_J$, placed in an external magnetic flux $\Phi_{ext}$. (b) Potential energy and energy levels of SPQC system for $L = 2L_J = 700\,pH$ and $\Phi_{dc} = 0.54\Phi_0$ versus $\varphi$ between zero and unity. The time-dependent flux amplitudes applied to the transitions $|0\rangle \to |1\rangle$, $|1\rangle \to |2\rangle$ and $|2\rangle \to |3\rangle$ are shown by $\Omega_{10}$, $\Omega_{21}$ and $\Omega_{32}$, respectively.

**Figure 2.** The real (a) and imaginary (b) parts of the susceptibility versus probe detuning for different values of the Rabi frequency of applied microwave fields. Used parameters are $\Gamma_{10} = \Gamma$, $\Gamma_{21} = 1.57\Gamma$, $\Gamma_{32} = \Gamma$, $\gamma_{10} = \Gamma$, $\gamma_{20} = 2.29\Gamma$, $\gamma_{21} = 2.57\Gamma$, $\gamma_{30} = 2.43\Gamma$, $\gamma_{31} = 1.71\Gamma$, $\gamma_{32} = \Gamma$, $\Delta_m = 0$, $\Delta_c = 0$, $\Omega_{10} = 0.01\Gamma$, $\Omega_{32} = \Gamma$, $\Omega_{21} = \Gamma$ (solid), $5\Gamma$ (dashed), $8\Gamma$ (dash-dotted).

**Figure 3.** The corresponding group index behaviors versus probe detuning corresponding to Fig. 2. Used parameters are same as in Fig. 2.

**Figure 4.** The population difference of probe transition versus the Rabi frequencies. Used parameters are same as in Fig. 2.

**Figure 5.** The real (a) and imaginary (b) parts of $\rho_{30}$ versus $\Delta = \Delta_{10} + \Delta_{21} + \Delta_{32}$ for $\Delta_{21} = \Delta_{32} = 0$, $\Omega_{10} = \Omega_{21} = \Omega_{32} = 2.1\Gamma$ (solid), $5\Gamma$ (dashed). Other parameters are same as in Fig. 2.

**Figure 6.** The population difference of states $|0\rangle$ and $|3\rangle$ versus Rabi frequency in zero detuning of the applied fields for $\Omega_{10} = \Omega_{21} = \Omega_{32} = \Omega$. Other used parameters are same as in Fig. 2.

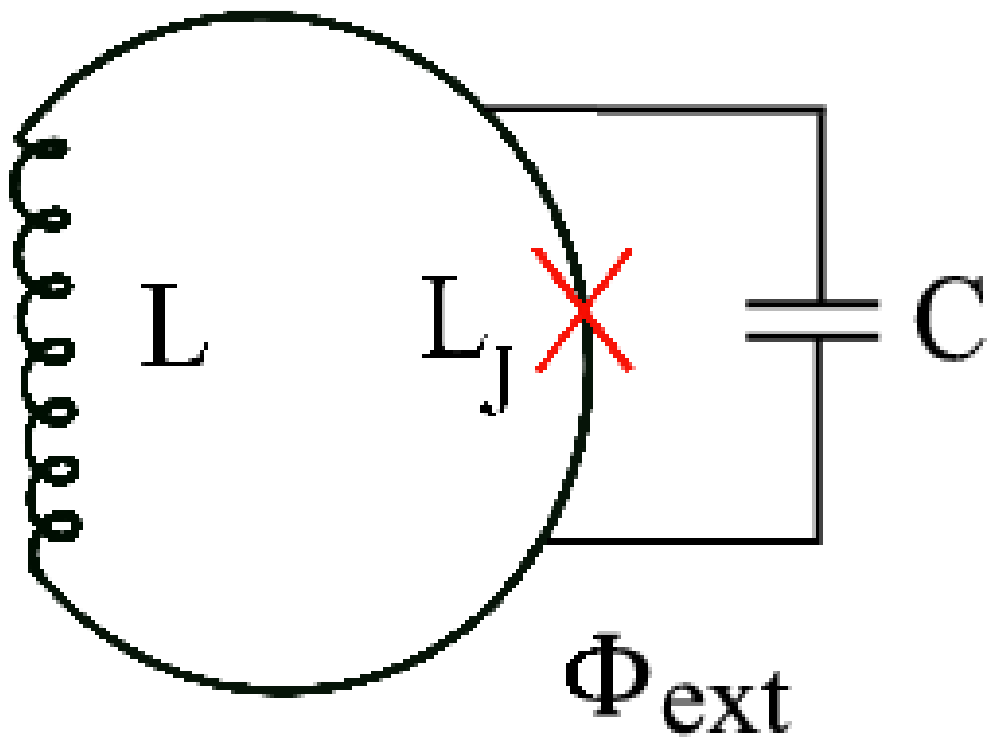

L
$L_J$
C
$\Phi_{ext}$

(a)

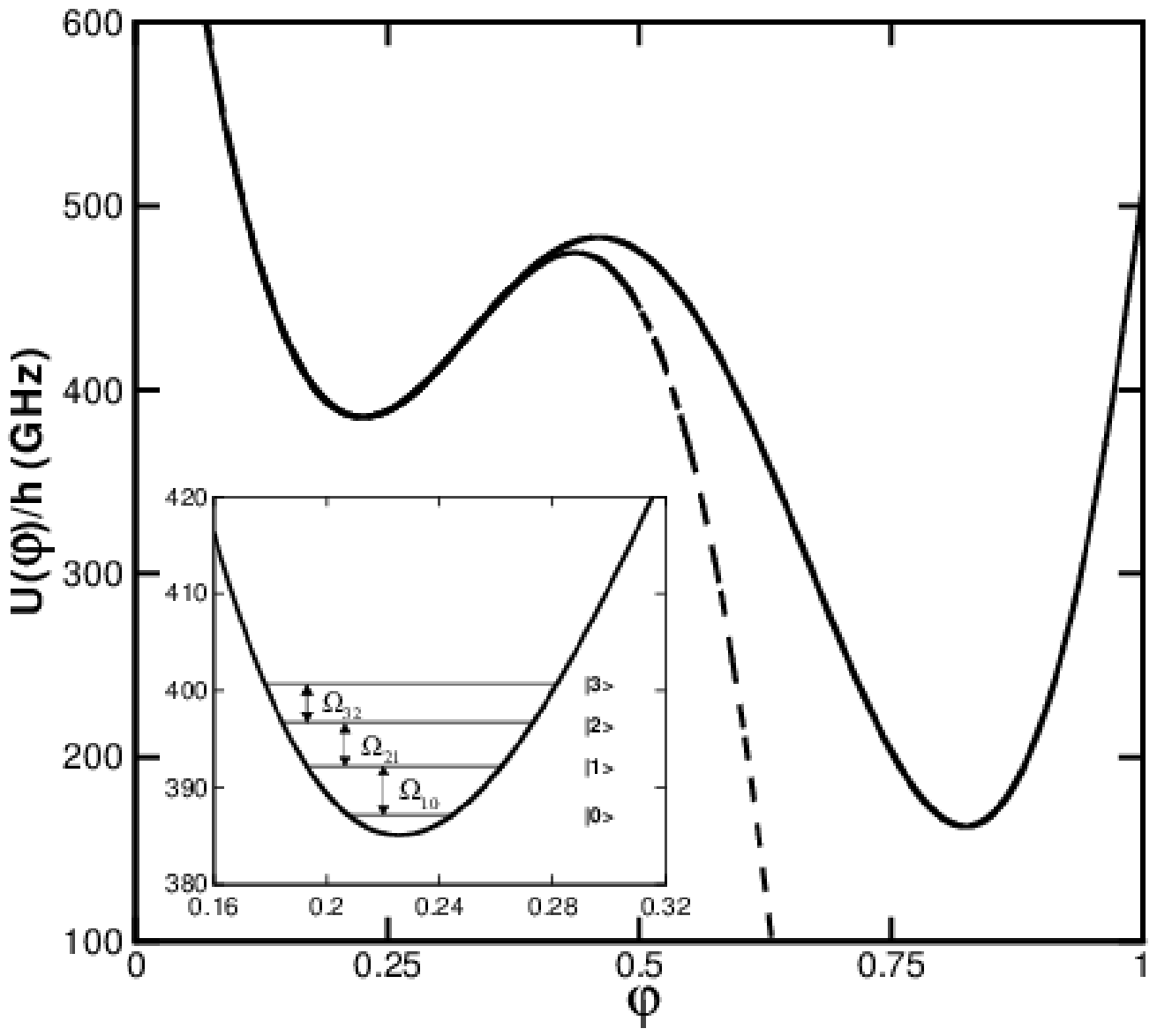

(b)

Figure 1

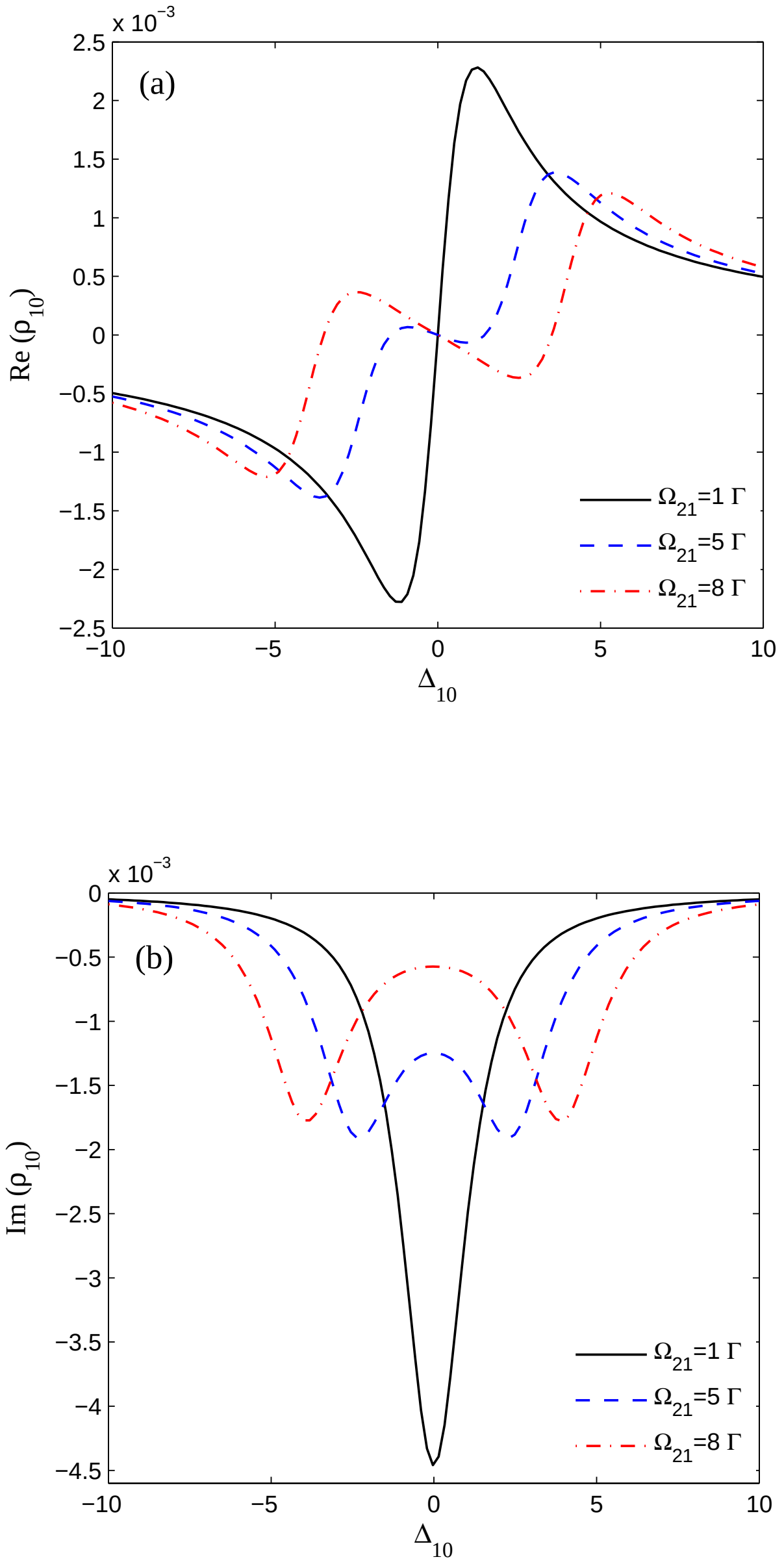

Figure 2

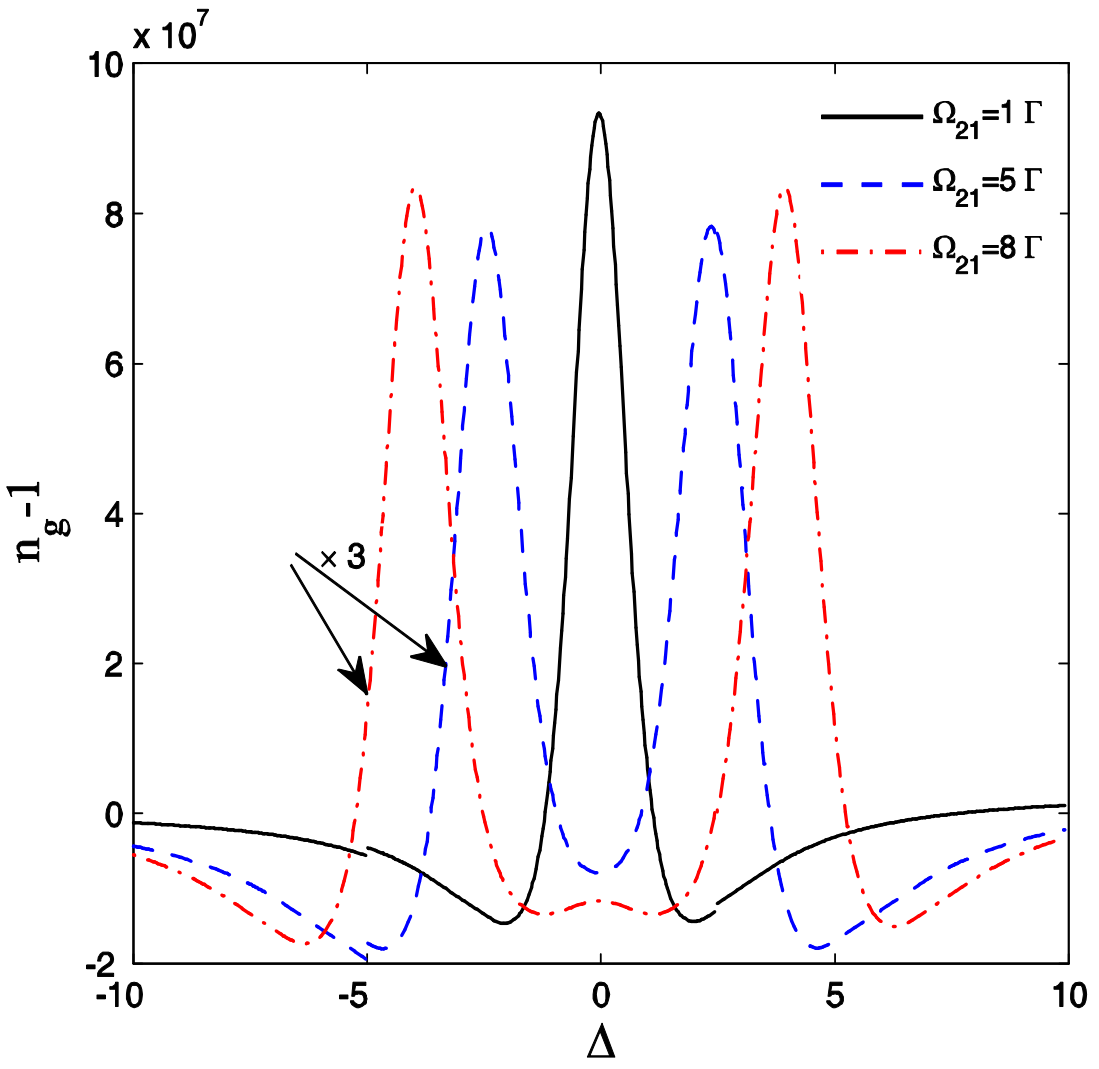

Figure 3

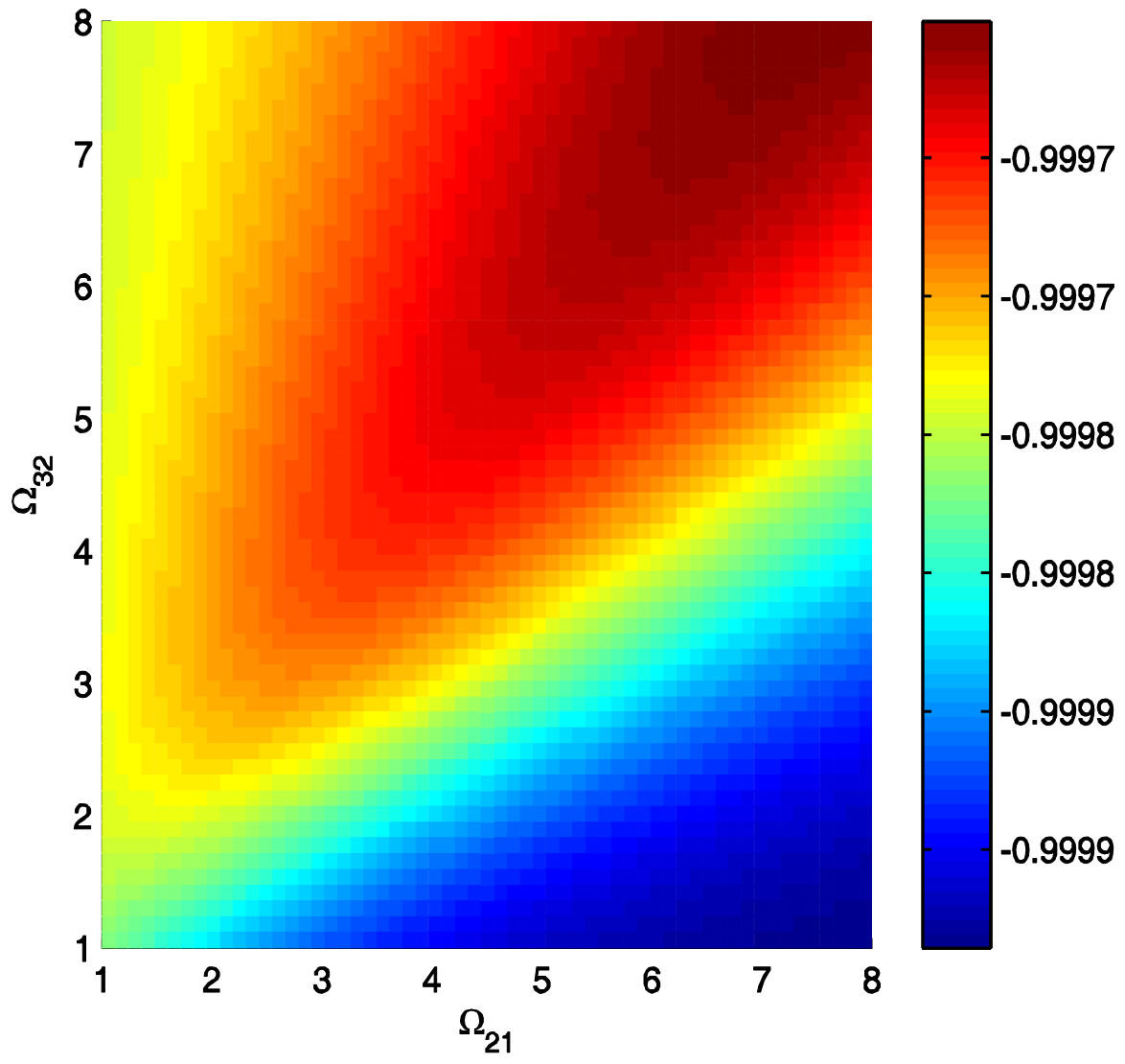

Figure 4

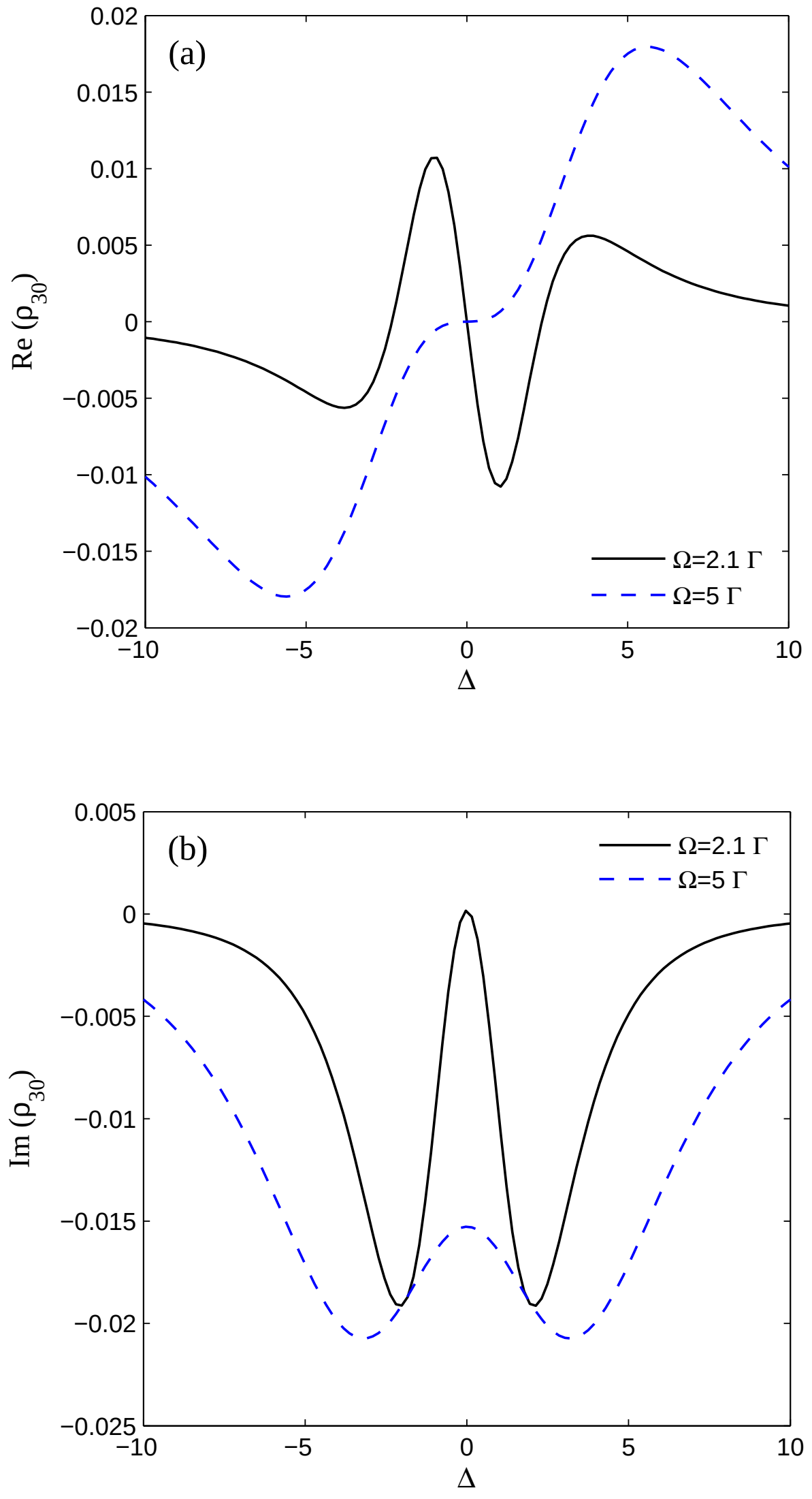

Figure 5

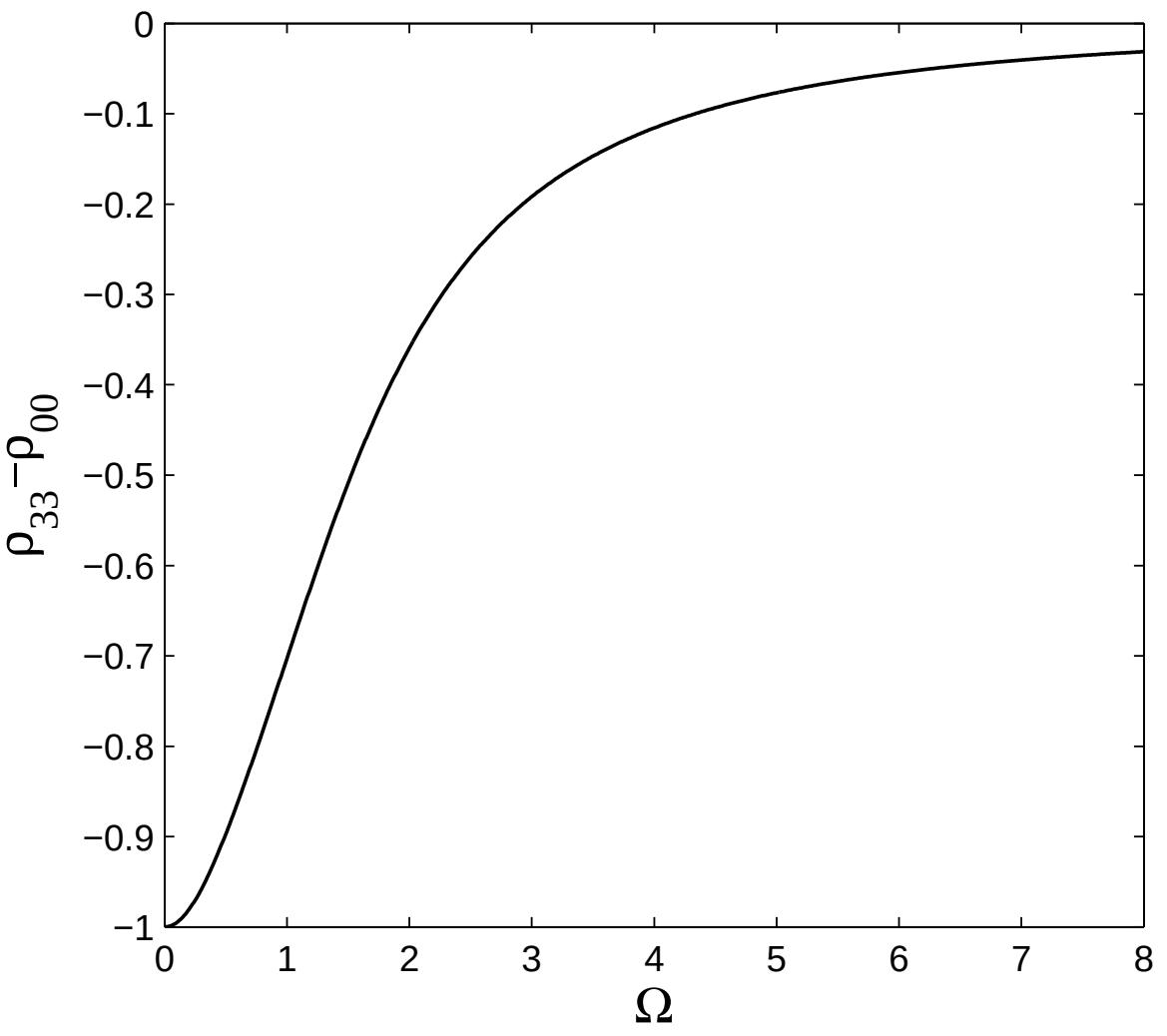

Figure 6